\documentstyle[preprint,aps,tighten] {revtex}
\begin{document}
\preprint{cond-mat/9402043}
\preprint{\begin{tabular}[t]{l}
T94/007 \\ cond-mat/9402043 \end{tabular}}
\draft

\title{Finite-lattice extrapolations for a\\
Haldane gap antiferromagnet}

\author{ O. Golinelli\thanks{
email: golinel, thierry and lacaze@amoco.saclay.cea.fr},
Th.\ Jolic\oe ur\thanks{CNRS Research Fellow.}
and R. Lacaze$^{\dag}$}

\address{Service de Physique Th\'eorique,
CEA Saclay\\ F-91191 Gif-sur-Yvette, France}

%\date{}
\maketitle
%%%%%%%%%%%%%%%%%%%%%%%%%%%%%%%%%%%%%%%%%%%%%%%%%%%%%%%%%%%%%%%%%%%%%%%%
\begin{abstract}
%%%%%%%%%%%%%%%%%%%%%%%%%%%%%%%%%%%%%%%%%%%%%%%%%%%%%%%%%%%%%%%%%%%%%%%%
We present results of exact diagonalizations of the isotropic
antiferromagnetic $S=1$ Heisenberg chain by the Lanczos method,
for finite rings of up to $N=22$ sites.
The Haldane gap $G(N)$ and the ground state energy per site $e(N)$
converge,
with increasing $N$, faster than a power law.
By VBS and Shanks transformations, the extrapolated values are
$G(\infty)= 0.41049 (2)$ and $e(\infty) = -1.401485 (2)$.
The spin-spin correlation function is well fit by
$\exp(-r/\xi)/\sqrt{r}$ with $\xi=6.2$.
\end{abstract}

\pacs{PACS number:  75.10.Jm}
\narrowtext
%%%%%%%%%%%%%%%%%%%%%%%%%%%%%%%%%%%%%%%%%%%%%%%%%%%%%%%%%%%%%%%%%%%%%%%%
\section{Introduction}
%%%%%%%%%%%%%%%%%%%%%%%%%%%%%%%%%%%%%%%%%%%%%%%%%%%%%%%%%%%%%%%%%%%%%%%%
A great variety of magnetic phenomena can be understood by
the study of classical spin systems. However, we know that there are
some surprises from quantum mechanics in one dimensional spin systems.
Haldane has conjectured \cite{Haldane}
that the properties of the one-dimensional Heisenberg
antiferromagnet are qualitatively different
depending on whether the spin
is integer or half-integer. This intriguing argument applies notably to
the simple prototypical antiferromagnetic (AF) $S=1$ spin chain which,
according to Haldane,
should exhibit a gap ($G$) towards spin excitations.
This conjecture has been checked experimentally, in particular with the
compound NENP, for which inelastic neutron scattering and
susceptibility measurements have clearly
shown the existence of a spin gap \cite{renard}.

The Heisenberg  AF $S=1$ spin chain has been studied in many numerical
works.
In 1973, ten years before the Haldane conjecture, De Neef \cite{deneef}
computed
the specific heat by exact diagonalizations of the Hamiltonian for
chain lengths up to $N=8$. In 1977, Bl\"ote \cite{blote} diagonalized
chains up to $N=10$. In 1982, Botet and Jullien \cite{botet}, with
diagonalizations up to $N=12$, obtained evidence for a non-vanishing gap
in the thermodynamic limit. Their value for the gap was rather imprecise
($G \approx 0.25J$). After this work, other authors used exact
diagonalizations with chains of increasing length:
in 1984, Glaus and Schneider \cite{glaus}
and Parkinson and Bonner \cite{parkinson}
with $N=14$, in 1987, Moreo \cite{moreo} with $N=16$
and in 1990, Lin \cite{lin} with $N=18$,
a length that has also been reached by Haas et al. \cite{haas} and
Takahashi \cite{takahashi}.
This growth is almost linear: two more spins every three years.
In fact, the numerical complexity of an exact diagonalization is $3^N$.
The exponential growth of computer power is not sufficient to explain
these results and much progress has been done in the algorithms.
However it is clear that it is very difficult
to continue along this line of study.

With Monte-Carlo methods \cite{takahashi,takahashi2,nomura,liang,goli},
longer chains can easily be studied (for example $N=64$),
but the results have statistical as well as systematic errors
(the latter being much more troublesome).
In 1985, Nightingale and Bl\"ote \cite{nightingale} obtained
a very precise estimate of the Haldane gap $G=0.41$
by use of a stochastic
implementation of direct iteration. In this case there is a systematic
bias,
caused by the finite number of "walkers", that is difficult to control.

Real-space renormalization-group methods have  also been applied to spin
chains.
The first works were quite disappointing because of large systematic
errors,
but in 1988 Lin and Pan \cite{linpan}, gave the very precise value
$G=0.4097(5)$.
In 1992, White and Huse \cite{white} improved this method and
found $G=0.41050(2)$ and a ground state energy with a precision of
$10^{-12}$.

In this work we have used the Lanczos technique on the longest possible
chain
we could reach which is $N=22$ spins and then we have applied
powerful extrapolation techniques. In this strategy, the only source
of error is due to the extrapolation technique while the finite-lattice
numbers are limited essentially by machine accuracy.

In Section II, we explain the numerical method (in particular
the importance of the symmetries and the quantum numbers of
the Haldane triplet).
The programming techniques useful for a chain length
$N=22$ are described in an Appendix.
In Section III, we explain our extrapolation method: the Shanks and VBS
transformations and how we quote errors.
In Section IV, we apply our strategy to the Haldane gap and the ground
state energy.
In Section V, we compare our results with those of other authors.
In particular, the precision for the gap is similar to that of
White and Huse \cite{white} with compatible results.
In Section VI, our results for the correlations functions are presented.
They are well described by a correlation length $\xi=6.2$.
Section VII presents our conclusions.

%%%%%%%%%%%%%%%%%%%%%%%%%%%%%%%%%%%%%%%%%%%%%%%%%%%%%%%%%%%%%%%%%%%%%%%%
\section{Numerical method}
%%%%%%%%%%%%%%%%%%%%%%%%%%%%%%%%%%%%%%%%%%%%%%%%%%%%%%%%%%%%%%%%%%%%%%%%
The Hamiltonian for a chain of length $N$ is:
\begin{equation} H = \sum_{i=1}^N \vec{S_i} \cdot \vec{S}_{i+1} \ ,
\end{equation}
where the $\vec{S_i}$ are the quantum spin-1 operators.
The exchange constant is positive ($J=1$) in the antiferromagnetic case.
The boundary conditions are periodic ($N$+1 $\equiv$ 1)
and the length $N$ is even to avoid frustration.

We have computed by exact diagonalization of the hermitian matrix
$H$, the ground state $|0\rangle$, its energy $E_0$, its
spin-spin correlation functions  and the energy of the first excitation
$E_1$,  for finite lengths up to $N=22$ spins.
The Haldane gap $G(N)$ is the difference between $E_0$ and $E_1$,
the two lowest eigenvalues.
The extrapolation method, which gives an estimation of $G(\infty)$ is
explained in the next section.
We have used the standard Lanczos method \cite{lanczos}, which is
well adapted for this problem: the matrix is very sparse and only a few
extreme eigenvalues are needed.

The size of the matrix $H$ is $3^N \times 3^N$.
With the symmetries of hamiltonian, $H$ is block diagonal and only
the two interesting blocks, with a size smaller than $3^N$, must be
diagonalized.
The symmetries of the lattice are the translational invariance $T$, and
the left-right reflection $Lr$
($Lr$ transforms the wave vector $k$ to $-k$;
so it reduces the size of blocks only for $k=0$ or $\pi$).
The spin symmetries are the global rotation, $\vec{S}= \sum_i \vec{S_i}$
(it seems difficult to implement this symmetry, and
in practice, only a sub-group of $SU(2)$ is used).
The matrix elements are computed in the $z$-axis basis:
$\{|s_1, s_2, \dots, s_N\rangle \}$
where $s_i=-1,0$ or 1, are the eigenvalues of $S_i^z$.
In this basis, $T|s_1, \dots, s_N\rangle  = |s_2,\dots,s_N,s_1\rangle$
and $Lr|\{s_i\}_i\rangle  = |\{s_{N-i}\}_i\rangle$.
The spin symmetries that can be implemented easily are: $S^z$,
the magnetization along the
$z$-axis (which is diagonal in the $z$-axis basis)
and the $\pi$ rotation around the $x$-axis, $R_x = \exp (i \pi S^x)$. In
this basis, $R_x |\{s_i\}_i\rangle  = |\{-s_i\}_i\rangle$ and
the action of $R_x$ is a flip of all the spins.
So, $R_x$ maps the subspace $S_z=m$ onto $S_z=-m$ and reduces the
size of blocks only for $m=0$.

We will now explain which blocks contain the ground state
$|0\rangle$, and the first excitation $|1\rangle$.
Because of the $SU(2)$ symmetry, the eigenvectors can be labeled by
the quantum numbers $j$ and $m$.
Each energy level has a degeneracy $2j+1$ and a representative member in
the subspace $m=0$.
On the other hand, the subspace $m=1$ contains no singlet $j=0$.
With the help of general arguments \cite{marshall},
the ground state $|0\rangle$ of an antiferromagnetic model is a
singlet $j=0$, but the first excitation can have $j=0$ or 1.
The full diagonalization of $H$ for short chains shows
that the first excitation has $j=1$: the Haldane triplet.
We denote it $|1,m\rangle$ with $m=-1,0$ or 1.
The other quantum numbers are obtained by using the
transformation \cite{marshall}
\begin{equation}
U = \exp\left(i \pi \sum_{j {\rm \ even}} S_j^z\right).
\end{equation}
which is diagonal in the $z$-axis basis.
The interest of $U$ is that the non-diagonal elements of
$UHU^{-1}$ are 0 or $-1$.
The Perron-Frobenius theorem \cite{gantmacher} can then be applied in
each subspace $S_z=m$.
For $m=0$ or 1, it follows that the components of $U|0\rangle$
and $U|1,1\rangle$ are
strictly positive (in the $z$-axis basis).
In a subspace $S_z=m$,
$T\cdot U = (-1)^m U\cdot T$. So $|0\rangle$ has the wave vector
$k=0$ and $|1,1\rangle$ (and so the entire Haldane triplet
$|1,m\rangle$) has $k=\pi$.
The left-right reflection $Lr$ commutes with $U$;
$|0\rangle$ and $|1,1\rangle$
(therefore all the triplet) are even for $Lr$.
The spin rotation $R_x$ commutes with $U$;
$|0\rangle$ is even for $R_x$.
But $R_x|1,1\rangle =\pm |1,-1\rangle$ and only $|1,0\rangle$
is a eigenvector of $R_x$.
In a triplet,
the eigenvalues of $S_x$ are $-1$,0 and 1; so those  of $R_x$ are
$-1$, 1 and $-1$.
The eigenvectors $|1,1\rangle  \pm |1,-1\rangle$ match $\pm 1$;
necessarily, the eigenvalue of $|1,0\rangle$ for $R_x$ is $-1$.
To summarize, the ground state $|0\rangle$ is in the subspace
$(S_z=0, k=0, Lr=+1, R_x=+1)$ and one representative of the Haldane
triplet is the ground state of the subspace $(S_z=0, k=\pi, Lr=+1,
R_x=-1)$.

The advantage of these symmetries is the reduction of the size of the
matrix.
The size of the subspace $S_z=m$ is
$\sum N!/(n_+!\:n_0!\:n_-!)$ with $n_++n_0+n_-=N$ and $n_+-n_-=m$.
When $N$ is large,
\begin{equation}
\dim(S_z=0) \sim {1\over 2} \sqrt{3\over\pi} {3^N\over\sqrt N} \ \cdotp
\end{equation}
The translation $T$ reduces this size by a factor $N$
(asymptotically when $N$ is large), the left-right reflection $Lr$ by 2
($N$ large) and the $\pi$-rotation $R_x$ by 2 ($N$ large).
For $N=22$, the size is reduced by 851 ($1\%$ better than the asymptotic
formula) and it is equivalent to $N=16$ without symmetries.

Certain methods are well adapted to
obtain the ground state of a large, unstructured
sparse and symmetric (or hermitian) matrix,
for example the conjugate gradient \cite{nightingale93} and the
Lanczos \cite{lanczos} method.
In these iterative methods,
by starting with an arbitrarily vector $V_0$,
the matrix $H$ acts only in matrix-vector multiplications and remains
sparse:
the vector $V_n$ is a linear combination of  $H\cdot V_{n-1}$ and
the previous $V_i$'s ($i\leq n-1$).
Then $V_n$ is in the subspace
${\cal K}_n = {\rm span}(V_0,H\cdot V_0,\dots,H^n\cdot V_0)$.  From
a theoretical point of view, the Lanczos method builds an orthogonal
basis of ${\cal K}_n$ and the projection of $H$ on ${\cal K}_n$ is a
tridiagonal matrix $n\times n$.
After $n$ iterations, the ground state is approximated
by the vector of ${\cal K}_n$ which minimizes
the Rayleigh quotient $R(V)= \langle V|H|V\rangle /\langle V|V\rangle$.
So, by construction, it is the fastest convergent method among these one
using $H\cdot V$ multiplications.
Because computers have a finite precision,
the orthogonality of the $V_i$
tends to be lost after many iterations and it is difficult to obtain
many eigenvalues.
However, as we want only the ground states of some blocks, we used the
Lanczos method. For $N=22$, only 55 iterations (or matrix-vector
multiplications) are needed to obtain eigenvalues with a precision which
can not be improved by more iterations.
Details on the programming techniques are given in Appendix.

%%%%%%%%%%%%%%%%%%%%%%%%%%%%%%%%%%%%%%%%%%%%%%%%%%%%%%%%%%%%%%%%%%%%%%%%
\section{Numerical results and extrapolation method}
%%%%%%%%%%%%%%%%%%%%%%%%%%%%%%%%%%%%%%%%%%%%%%%%%%%%%%%%%%%%%%%%%%%%%%%%
The numerical values are given in Table \ref{results} with 12 digits
after the decimal point, for periodic chains with length up to $N=22$.
We have no direct means to estimate the precision.
Some direct iterations have been made with the eigenvector obtained by
the Lanczos method and the precision is estimated at better
than $10^{-11}$
for the ground state energy $E_0$ and the first excitation $E_1$.
The sizes $N=2$ and 4 are added because we will see that they have,
surprisingly, a good behavior with respect to
the extrapolation method. Results up
to $N=18$ have been published by other authors (see the caption of
Table \ref{results}).

The gap $G(N)=E_1-E_0$ and the ground state energy per site $e(N)=E_0/N$
have still not converged.
To obtain a good estimate of their thermodynamic limits, the
convergence must be accelerated by an extrapolation method, adapted to
their behavior.

For periodic chains, the convergence of the gap has been observed to be
exponential.
In a theory with a gap, we expect of course exponential convergence
towards the thermodynamic limit
for a closed chain. This has been shown explicitly in the large-N limit
of the nonlinear sigma model \cite{goli}.
We analyze the estimated decay length $\xi(N)$ at the index $N$
as given by
\begin{equation}
\xi(N) = 2 / \ln \left( {a_{N-4}-a_{N-2}} \over {a_{N-2}-a_N} \right ),
\label{xidef}
\end{equation}
where $a_N$ represents the sequence $G(N)$ or $e(N)$.
If $a_N = A + b\, \exp(-N/\xi)$, then $\xi(N)=\xi$ exactly for all
$N$.
If $a_N= A + b/(N-n_0)^\nu$, then $\xi(N)\sim (N-n_0)/(\nu+1)$ for $N$
large, and the exponent $\nu$ is given by the asymptotic slope of
$\xi(N)$.

In Fig.\ \ref{xi}, we have plotted $\xi(N)$ for $G(N)$ and $e(N)$.
Both curves are concave:
the estimated exponent $\nu$ increases with $N$
(for $N=22$, respectively, $\nu\approx 15$ and 11).
This figure shows that the gap and the energy per site converge faster
than a power law. This is good evidence for the expected exponential
behavior of the Haldane conjecture.
For this reason,
we extrapolate with the Shanks transformation \cite{shanks,brez}.
We explain in some detail this method because we will use it in
a different way than Ref.\ \onlinecite{goli} or \onlinecite{sakai}.
If the sequence $a_N$ is a sum of $k$ exponentials:
\begin{equation}
a_N = A + b_1\,e^{-N/\xi_1} + \dots + b_k\,e^{-N/\xi_k},
\label{expo}
\end{equation}
the limit $A$ is one of the $2k+1$ unknowns and
can be obtained by solving the system (\ref{expo}) for
$a_{N-2k},\dots,a_N,a_{N+2},\dots,a_{N+2k}$.
We call $A_N^{(k)}$ this solution, i.e.\ the limit $A$ extracted from
$a_{N-2k},\dots,a_{N+2k}$ supposing that the sequence $a_N$ is a sum
of $k$ exponentials.
Of course, if the sequence $a_N$ has not exactly this form, then
$A_N^{(k)}$ varies with $N$.
The simplest case of the Shanks transformation is $k=1$.
The solution is
\begin{equation}
A_N^{(1)} = {a_{N+2}\,a_{N-2}-a_N^2 \over a_{N+2} - 2a_N + a_{N-2}},
\label{k1}
\end{equation}
which is also called Aitken's $\Delta^2$ process.
This Aitken-Shanks transformation (\ref{k1}) with $k=1$
can be iterated \cite{goli,brez,sakai}:
it is first applied to $a_N$, then again to the obtained sequence
$A_N^{(1)}$, and so forth. The iteration of the
($k$=1)-transformation does not give the $A_N^{(k)}$, for $k>1$.
For $k\geq 1$, the $A_N^{(k)}$ can be computed by using the
recursive ``cross rule'' due to Wynn \cite{brez,wynn}
\begin{equation}
(A_N^{(k+1)} - A_N^{(k)})^{-1} + \alpha
(A_N^{(k-1)} - A_N^{(k)})^{-1} =
(A_{N-2}^{(k)} - A_N^{(k)})^{-1} +
(A_{N+2}^{(k)} - A_N^{(k)})^{-1} ,
\label{cross}
\end{equation}
with the initial conditions $A_N^{(0)}=a_N$, $A_N^{(-1)}=\infty$ and
where $\alpha=1$ (for the Shanks transformation).
The table of the
$A_N^{(k)}$ verify many algebraic properties \cite{brez}; for example,
if the $a_N$ are the partial sums of a power series,
the $A_N^{(k)}$ are the Pad\'e approximants of the series.
Due to the non-linearity of Eq.\ (\ref{cross}), only a few
exact results are known.
For example, if $a_N$ is a sum of exponentials
\begin{equation}
a_N \sim A + \sum_{i=1}^{\infty} b_i\,e^{-N/\xi_i}
\quad {\rm when} \quad  N \rightarrow \infty
\end{equation}
with $\xi_1>\xi_2>\cdots >0$,
then \cite{brez} for $k$ fixed and $N \rightarrow \infty$
\begin{equation}
A_N^{(k)}\sim A+b_{k+1}{(\lambda_{k+1}-\lambda_1)^2\cdots
(\lambda_{k+1}-\lambda_k)^2 \over
\lambda_{k+1}^k(1-\lambda_1)^2\cdots (1-\lambda_k)^2} e^{-N/\xi_{k+1}}
\quad{\rm with}\quad\lambda_i=e^{-2/\xi_i},
\label{asymp}
\end{equation}
Each $k$-iteration removes a exponential and
each column $k$ will be more rapidly convergent
than the previous ones when $N$ goes to the infinity.
Even if the exact results are rare, in practice a quite general class
of sequence is accelerated.

The parameter $\alpha$ in Eq.\ (\ref{cross}) was introduced by Vanden
Broeck and Schwartz \cite{vbs} (VBS).
The table changes when $\alpha$ varies.
The Pad\'e-Shanks transformation (at order $k$) is given by $\alpha=1$.
The iterated Aitken-Shanks ($k$=1)-transformation is
given by $\alpha=0$.
When $\alpha=-1$, Hamer and Barber \cite{hamer} have shown that,
if $a_N$ has exactly the power law behavior
$a_N= A + b (1-\nu/2)(1-{\nu/4})\dots(1-{\nu /N})$
(for $N$ large, $a_N \sim A+b'/N^\nu$),
the second column $A_N^{(2)}=A$ for all $N$.
Hamer and Barber's transformation can be iterated with
$\alpha_k=0,-1,0,-1,\dots$ successively for each column $k$.

%%%%%%%%%%%%%%%%%%%%%%%%%%%%%%%%%%%%%%%%%%%%%%%%%%%%%%%%%%%%%%%%%%%%%%%%
\section{Extrapolated values}
%%%%%%%%%%%%%%%%%%%%%%%%%%%%%%%%%%%%%%%%%%%%%%%%%%%%%%%%%%%%%%%%%%%%%%%%
On Table \ref{gap},
we give the results of the Shanks transformation ($\alpha=1$ and $k=1$
to 5) for the gap $G(N)=E_1-E_0$.
The estimated decay lengths $\xi(N,k)$, defined by Eq.\ (\ref{xidef})
are calculated, for each $k$, with the $A_N^{(k-1)}$.
One remarks that the data of each column are monotonic.
This is also true with the last oblique row.
In particular, the $\xi(N,k)$ decrease with $k$. In fact because of the
rather small values of $N$, $\xi(N,k)$ represent only an effective
decay length and $\xi(N,k)$ involve data from smaller value of $N$ than
$\xi(N,k-1)$. Thus what we have to expect is
\begin{equation}
\xi(N,k) < \xi(N+2,k-1) .
\label{chikn}
\end{equation}
These inequalities are all satisfied and
the idea that the Shanks transformation removes at
each step an exponential transient is coherent.
The values for $N=2$ and 4 do not disturb the table.
One must compute this table with at least double precision arithmetic
(64 bits), because large cancellations occur in the cross rule
(\ref{cross}). We verified that the precision is better than
$10^{-7}$ (only 6 digits are retained).

If the parameter $\alpha$ of Eq.\ (\ref{cross}) takes a value
sufficiently different from $\alpha=1$, this table loses its consistency
(for example, Eq.\ (\ref{chikn}) is not verified for all ($N,k$) values)
and the extrapolation cannot be trusted.
In particular, $\alpha=0$ (iterated Aitken-Shanks transformation) and
$\alpha_k=0,-1,0,-1,\dots$ (Hamer and Barber's transformation) do
not give a satisfactory table.
For the gap, the acceptable interval is $0.7<\alpha<1.05$, (close to
$\alpha=1$, the pure Shanks transformation) and the
respective extrapolations $A^{(5)}$ for these bounds are
$0.410478$ and $0.410504$.
Then, with the VBS or  Shanks transformation,
for periodic chains with $N$ up to 22, the gap value is
\begin{equation}
G(\infty)= 0.41049 (2).
\label{g}
\end{equation}
The error bar ($2.10^{-5}$) is an estimation of the systematic error
due to the arbitrary parameter $\alpha$.
The Shanks transformation does not give a direct estimation of
the systematic error with respect to the true limit.
But the main argument in favor of this method is the good
regularity of the data of the Table \ref{gap}.

Now we study the ground state energy per site $e(N)=E_0/N$ with a same
method.
In Table \ref{energy},
we give the results of the Shanks transformation ($\alpha=1$ and $k=1$
to 5).
The convergence is faster (in particular, the decay lengths $\xi$ are
shorter) and the precision is better ($10^{-9}$ but only 6 digits are
given here). For $\alpha=1$ (the pure Shanks transformation),
the table is consistent: all the columns are monotonic,
as is the last oblique row, and the $\xi$'s decrease with $k$.
For $\alpha$, the acceptable interval is $0.5<\alpha<1.1$.
The respective extrapolation $A^{(5)}$ for these bounds are
$-1.401484$ and $-1.401487$.
Then, with the VBS or Shanks transformation, for periodic chains with
$N$ up to 22, the ground state energy per site is
\begin{equation} e(\infty) = -1.401485 (2).  \label{e} \end{equation}

%%%%%%%%%%%%%%%%%%%%%%%%%%%%%%%%%%%%%%%%%%%%%%%%%%%%%%%%%%%%%%%%%%%%%%%%
\section{Comparisons}
%%%%%%%%%%%%%%%%%%%%%%%%%%%%%%%%%%%%%%%%%%%%%%%%%%%%%%%%%%%%%%%%%%%%%%%%
We compare our extrapolations with those published by other authors.
For the most part, numerical results have been obtained by exact
diagonalizations (with simple iterations or Lanczos method), Monte-Carlo
simulations or the real-space renormalization-group method.
Of course, for a fixed chain length $N$, we find the same results as the
other exact diagonalization studies:
Bl\"ote \cite{blote} for $N=10$, Glaus and Schneider \cite{glaus} and
Parkinson and Bonner \cite{parkinson} for $N=14$, Moreo \cite{moreo} for
$N=16$ and Lin \cite{lin} for $N=18$.
Sakai and Takahashi \cite{sakai}, with $N\leq 16$ results extrapolated
by the Aitken-Shanks iterated transformation (Eq.\ \ref{k1}) give
$G(\infty)=0.411 (1)$, compatible with Eq.\ (\ref{g}).

The results of Monte-Carlo calculations have statistical as well as
systematic errors but longer chains (for example $N=32$ or 64) can be
studied.
The stochastic iteration of Nightingale and Bl\"ote \cite{nightingale}
gives compatible results: $G(32)=0.413 (7)$ and $e(32)=-1.40155 (16)$,
as the method of Liang \cite{liang}: $e(64)=-1.402 (1)$.
On the other hand, the projector Monte-Carlo method of Takahashi
\cite{takahashi} is not compatible: $e(32)=-1.4023 (1)$.
The methods based of the Trotter-Suzuki decomposition are characterized
by imprecisions when the temperature goes to zero and
the properties of the ground state are not well reproduced.
For example Nomura \cite{nomura} gives $G=0.425$.

Concerning the real-space renormalization-group method
(or truncated basis expansion), we are in disagreement with some of the
published values:
Pan and Chen \cite{pan} ($G=0.368166$ and $e=-1.449724$),
Mattis and Pan \cite{mattis} ($e=-1.388$), and
Xiang and Gehring \cite{xiang} ($e(\infty)=-1.377$).
On the other hand, we aggre with Lin and Pan
\cite{linpan}: $e(\infty) = -1.4021 (5)$ and $G(\infty)=0.4097 (5)$
and with the recent method of White \cite{white} which gives
$e(\infty)=-1.401484038971(4)$ and $G(\infty)=0.41050 (2)$.
These values are in good agreement with ours; the ground state energy
is more precise and the precision on the gap value is similar.
The fact that both methods give results with 6 and 5 identical digits
is a good argument that there are both quite accurate.

%%%%%%%%%%%%%%%%%%%%%%%%%%%%%%%%%%%%%%%%%%%%%%%%%%%%%%%%%%%%%%%%%%%%%%%%
\section{Correlation functions}
%%%%%%%%%%%%%%%%%%%%%%%%%%%%%%%%%%%%%%%%%%%%%%%%%%%%%%%%%%%%%%%%%%%%%%%%
In this section, we present our results for the correlation functions
\begin{equation}
C_N(r)= (-1)^r\langle S_0^z S_r^z\rangle  =
(-1)^r\langle \vec S\vec S\rangle /3 \ ,
\end{equation}
for the ground state of a isotropic and
periodic chain of length $N$ for $N\leq 22$.
Numerical values are given in Table \ref{corr}.
To compute these quantities, the eigenvalue is not sufficient and the
eigenvector is required.
So the precision for the $C_N(r)$ is less than for the energies.
It can be estimated around $10^{-8}$, for example by
comparing  $C_N(1)$ with $E(N)$, or
by direct applications of the matrix $H$ on the Lanczos result.

For an infinite chain, the Haldane conjecture predicts an
exponential decrease
$C_\infty (r) \sim b \exp(-r/\xi)/\sqrt{r}$
when $r$ is large. This is because the underlying
continuum theory is a nonlinear sigma model which is {\it relativistic}
in 1+1 dimensions.
In fact, if we approximate the nonlinear sigma model by free massive
bosons, the propagator is the modified Bessel function $K_0$ which
has this asymptotic behavior.
The Haldane conjecture does not deal with short-distance details so
that there is no {\it a priori} preferred choice when trying to
fit data on the full range of spin-spin separation.

For a periodic chain, one has the equality $C_N(r)=C_N(N-r)$.
To extract the correlation length $\xi$ for $N$ and $r$ large, we
analyze our data with the guess
$C_N(r) = C_{\infty}(r) + C_{\infty}(N-r)$,
which is reasonable if $\xi \ll N$.
For some classical spin systems (one-dimensional Potts model, Ising
chain with a magnetic field, \dots) with non-vanishing temperature,
the corrections to this formula are of order ${\cal O}(C_{\infty}(N))$.

We have verified that $\exp(-r/\xi)/\sqrt{r}$ fits  the data better
than $\exp(-r/\xi)$ or $\exp(-r/\xi)/r$. From
these three forms, the estimated values for $\xi$ are respectively
6.2, 4.5 and 10, for $N=22$.
In Fig.\ \ref{bessel},
we compare the Bessel function $K_0(r/\xi)$ and $\exp(-r/\xi)/\sqrt{r}$.
Both fits are comparable in quality.
But the estimated correlation lengths $\xi$ are slightly different:
5.9 versus 6.2 for $N=22$.
For $N=10$, they are respectively 5.4 and 6.2.
We notice that the optimal $\xi$ is 6.2, for all $N\leq 22$, with
$\exp(-r/\xi)/\sqrt{r}$, whereas the estimation for $\xi$ with
the Bessel function vary with $N$.
For this reason, we prefer the former but we keep in mind that both laws
have the same asymptotic behavior and that only a exact solution of the
model can give the correlations for short range.

It is interesting to compare the correlation length $\xi$
(obtained with $C_N(r)$) and the decay length $\xi(N)$
(Eq.\ \ref{xidef}) of the gap $G(N)$ (Table\ \ref{gap})
and energy $e(N)$ (Table\ \ref{energy}).
On Fig.\ \ref{xi}, we see that it is not excluded that $\xi=6.2$ is
the limit of $\xi(N)$.
The extrapolation of $\xi(N)$ with the Shanks transformation gives
$5.5$ for the gap and $4.6$ for the energy, but
the columns of the Shanks table are non-monotonic and
these results are only qualitative.
Of course, by analyzing the convergence of the $G(N)$ and $e(N)$ with
an exponential corrected by a power law (as $1/\sqrt N$),
the estimations
of $\xi(N)$ are greater, and the extrapolations are closer from $6.2$.
Then this comparison is only qualitative and requires longer chains.
Our value ($\xi=6.2$)
is equal to the estimate quoted by Nomura \cite{nomura}
($\xi=6.25$) and by Liang \cite{liang} ($\xi=6.2$) with Monte Carlo
methods for $N=64$.
It is comparable with the results of Takahashi \cite{takahashi2}
($\xi=5.5 \pm 2.$)  with a Monte Carlo method for $N=64$,
by Kubo \cite{kubo} ($\xi=6.7$) with a transfer-matrix method and
by White and Huse \cite{white} ($\xi=6.03$) with the real-space
renormalization-group method.

%%%%%%%%%%%%%%%%%%%%%%%%%%%%%%%%%%%%%%%%%%%%%%%%%%%%%%%%%%%%%%%%%%%%%%%%
\section{Conclusion}
%%%%%%%%%%%%%%%%%%%%%%%%%%%%%%%%%%%%%%%%%%%%%%%%%%%%%%%%%%%%%%%%%%%%%%%%
The main limitation of exact
diagonalizations is, of course, the small lengths that can be studied.
The numerical complexity grows as the exponential $(S(S+1))^N$ for $N$
spins $S$ and the limits of computer power are fastly reached.
The length $N$ of the system must be compared with the physical
correlation length $\xi$, and in fact, the situation for the $S$=1 AF
spin chain is quite favorable. Within the Haldane conjecture, $\xi$ is
finite for integer  spins and shortest for small $S$.
We have shown that some quantities can be measured with excellent
accuracy: the gap and the ground state energy.
On the other hand,  the correlations
$C_{\infty}(r)$ (and thus the correlation length $\xi$) clearly need
longer chains.

The main advantages of exact diagonalizations are that they depend
only on one parameter (the size $N$) and give exact
results (i.e. with machine precision).
One has to deal only with the thermodynamic limit.
By comparison, methods based on Trotter-Suzuki decomposition have three
parameters (number of slices, temperature and length of the chain) and
systematic errors which decrease by extrapolating in the number of
slices.
Monte-Carlo methods have their own parameters (number of walkers or
length of simulations, \dots) which must be tuned, and the results have
statistical fluctuations as well as systematic errors.
Real-space renormalization-group methods have to extrapolate w.r.t. the
number of basis states and the chain length.

The high precision allows the use of sophisticated extrapolation methods
and we are able to validate some  assumptions on the asymptotic
behavior.
Fig.\ \ref{xi} suggests that the use of the Shanks transformation
is optimal concerning gap extrapolation. In fact the  parameter $\alpha$
of the more general VBS transformation can vary only in a small interval
around $\alpha=1$.
This shows that our choice is not arbitrary but dictated by the data.
The results of exact diagonalizations combined with a
careful extrapolation can give physical quantities in the thermodynamic
limit with a good precision.

%%%%%%%%%%%%%%%%%%%%%%%%%%%%%%%%%%%%%%%%%%%%%%%%%%%%%%%%%%%%%%%%%%%%%%%%
\appendix \section*{programming techniques}
%%%%%%%%%%%%%%%%%%%%%%%%%%%%%%%%%%%%%%%%%%%%%%%%%%%%%%%%%%%%%%%%%%%%%%%%
We used a Cray~2 of the CEA  with a central memory of 256 Megawords
of 64 bits.
Some details of our program are useful only for this kind of machine in
particular and are not described here.
The algorithm has two main parts, the building of the sparse matrix $H$
(and its storage on disks) and the matrix-vector $H\cdot V$
multiplication, needed for the Lanczos iterations.

We consider first the matrix multiplication.
The matrix  $H$ is very sparse. For $N=22$, its order is $\approx$ 37
million
and the number of non-vanishing elements is (on average) $8N/9$ per
rows (when $N$ is large).
We use the classical storage by rows with
only the non-vanishing elements of $H$ (values and column
number) stored.
In practice, one bit is needed for the value ($\pm 1$) and 26 bits for
the column number. So, two elements are stored in a 64-bits word and
$3.5$ Gigabytes are used for $H$.
The matrix-vector multiplication is done by an indirect addressing of
the elements of the vector, where the address is the column number.
This indirect addressing is the most time consuming part of the program
and it is intrinsic to this sparse storage method.
For $N=22$, a multiplication needs 190 seconds of one Cray~2 CPU.

The most difficult part of the algorithm is the building of the
matrix with use of  symmetries.
Each $z$-axis basis state $|s_1,\dots,s_N\rangle$ is described by the
number $\sum_r (s_r+1) 3^{r-1}$.
First, the list of the symmetrized basis states is obtained.
Each symmetrized state is represented by the state of the $z$-axis
basis, which contributes and has the smallest number.
Then the hamiltonian $H$ operates on this list and generates other
states.
The problem is to find the corresponding symmetrized states (and their
phases).
Possible methods are a) each generated state is symmetrized
by action of all the symmetries operators or b) a storage table gives,
for each $z$-axis state, the symmetrized one and the generated state is
searched in this table.
The first method is too time consuming and the second one uses too
much memory.

We use an intermediate method with a decomposition in two sublattices
\cite{lin}, $A=\{s_{2r}\}$ and $B=\{s_{2r+1}\}$.
The symmetries $R_x$, $Lr$ and $T^{2k}$ do not exchange $A$ and $B$.
We call them sublattice symmetries.
On the other hand, symmetries $T^{2k+1}$ exchange $A$ and $B$.
Then, each symmetry is the product of a sublattice symmetry and
possibly $T$.
Since a sublattice is described only by $3^{N/2}$ states, we can use a
storage table
which gives for each sublattice state the symmetrized one.
For all the chain, $A$-symmetrization consist to symmetrized $A$
and to operate on $B$ with the same operator.
Since a storage table of size $3^{N/2}$ is available, it does not
require much time or memory.
The last step is the action of $T$, which exchange $A$ and $B$,
and $S_z$, which imposes $S_z(A) + S_z(B) = 0$.
By symmetrizing by $T$, the number of the $A$-symmetrized states
(around 78 millions for $N=22$ and $S_z=0$) is
divided by two (for $N$ large).
On our computer, we keep on memory the list which gives the fully
symmetrized state for each $A$-symmetrized one.
This list has some properties of factorization, as well explained in
Ref.\ \onlinecite{lin}, and the location of each state is easily
obtained by considering each sublattice.
To summarize, a generated state is, in a first step, symmetrized by
$R_x$, $Lr$ and $T^{2k}$ (which let invariant the sublattices),
and in a last step by $T$.
The first step needs only short lists ($3^{N/2}$) and the final one
a big list, for which the length is two times the order of the fully
symmetrized block.
For $N=22$, our program needs 2200 seconds of one Cray~2 CPU  to built
the matrix $H$.

For one block ($|0\rangle$ or $|1,0\rangle$), with the Lanczos method,
the precision cannot be improved after 55 iterations (for $N=22$).
To compute eigenvectors, the Lanczos method is not optimal because all
the intermediate vectors must be stored.
To do that, 16 Gigabytes are required.
Then, a first Lanczos calculation gives the eigenvalues and the
coordinates of eigenvectors on the Lanczos basis.
A second Lanczos calculation is needed to generate the eigenvectors.

The computations of this article have used 12 hours of one Cray~2 CPU.

%%%%%%%%%%%%%%%%%%%%%%%%%%%%%%%%%%%%%%%%%%%%%%%%%%%%%%%%%%%%%%%%%%%%%%%%

%%%%%%%%%%%%%%%%%%%%%%%%%%%%%%%%%%%%%%%%%%%%%%%%%%%%%%%%%%%%%%%%%%%%%%%%
% figures
%%%%%%%%%%%%%%%%%%%%%%%%%%%%%%%%%%%%%%%%%%%%%%%%%%%%%%%%%%%%%%%%%%%%%%%%

\begin{figure}
\caption{The estimated decay length $\xi(N)$ given by
Eq.\ (\protect\ref{xidef}) for the gap $G(N)$ and the ground state
energy per site $e(N)$ vs.\ the length $N$ of the periodic chain.}
\label{xi}
\end{figure}

\begin{figure}
\caption{The ratio of the correlation function
$C_N(r)= (-1)^r\langle S_0^z\cdot S_r^z\rangle$
and two proposed laws versus $r$.
The $C_N(r)$ have been exactly computed for $N=22$.
The ratios are normalized to 1 for $r=N/2$.}
\label{bessel}
\end{figure}

%%%%%%%%%%%%%%%%%%%%%%%%%%%%%%%%%%%%%%%%%%%%%%%%%%%%%%%%%%%%%%%%%%%%%%%%
% tables
%%%%%%%%%%%%%%%%%%%%%%%%%%%%%%%%%%%%%%%%%%%%%%%%%%%%%%%%%%%%%%%%%%%%%%%%

\begin{table}
\caption{Dimension of the largest block $(S_z=0, k=0, Lr=+1, R_x=+1)$,
ground state energy $E_0$, first excitation energy $E_1$,
gap $G(N)=E_1 - E_0$
and ground state energy per site $e(N)=E_0/N$ for chain lengths $N=2$
to $22$. These results are obtained by exact diagonalization.
Previous results for $N=8$ are given by De Neef \protect\cite{deneef},
$N=10$ by Bl\"ote \protect\cite{blote},
$N=12$ by Botet and Jullien \protect\cite{botet},
$N=14$ by Glaus and Schneider \protect\cite{glaus} and Parkinson and
Bonner \protect\cite{parkinson},
$N=16$ by Moreo \protect\cite{moreo}
and $N=18$ by Lin \protect\cite{lin}.}
\label{results}
\begin{tabular}{rrrrrr}
$N$ & dimension & $-E_0$ & $-E_1$ & gap $G(N)$ & $-e(N)$\\
\hline
2  &        2   &  4.0             &  2.0            &
2.0            & 2.0           \\
4  &        5   &  6.0             &  5.0            &
1.0            & 1.5           \\
6  &       15   &  8.617423181814  &  7.896795819190 &
0.720627362624 & 1.436237196969\\
8  &       59   & 11.336956077897  & 10.743400823522 &
0.593555254375 & 1.417119509737\\
10 &      290   & 14.094129954932  & 13.569322004518 &
0.524807950414 & 1.409412995493\\
12 &     1 728  & 16.869556139477  & 16.385359669563 &
0.484196469914 & 1.405796344956\\
14 &    11 549  & 19.655133499935  & 19.196168152997 &
0.458965346938 & 1.403938107138\\
16 &    82 790  & 22.446807281171  & 22.004011719811 &
0.442795561360 & 1.402925455073\\
18 &   617 898  & 25.242312007671  & 24.810090537803 &
0.432221469868 & 1.402350667093\\
20 &  4 730 966 & 28.040291720480  & 27.615081406019 &
0.425210314461 & 1.402014586024\\
22 & 36 871 567 & 30.839898879910  & 30.419383859516 &
0.420515020394 & 1.401813585450\\
\end{tabular}
\end{table}

\begin{table}
\caption{The Shanks extrapolations $A_N^{(k)}$ for the gap values
$G(N)=E_1-E_0$ with $k=1$ to 5 (with the VBS parameter $\alpha=1$).
The estimated decay lengths $\xi$ are obtained by applying the formula
(\protect\ref{xidef}) on the $A^{(k-1)}$.}
\label{gap}
\begin{tabular}{r|c|cc|cc|cc|cc|cc}
$N$ & $G(N)$ & $A^{(1)}$ & $\xi$ &$A^{(2)}$ & $\xi$ &$A^{(3)}$ & $\xi$
&$A^{(4)}$ & $\xi$ &$A^{(5)}$ & $\xi$ \\
\hline
 2 & 2.000000 &          &      &          &      &          &      &
          &      &          &      \\
 4 & 1.000000 & 0.612320 & 1.57 &          &      &          &      &
          &      &          &      \\
 6 & 0.720627 & 0.487533 & 2.54 & 0.435259 & 1.91 &          &      &
          &      &          &      \\
 8 & 0.593555 & 0.443776 & 3.26 & 0.417985 & 2.28 & 0.412584 & 1.59 &
          &      &          &      \\
10 & 0.524808 & 0.425578 & 3.80 & 0.413089 & 2.43 & 0.411146 & 1.75 &
 0.410712 & 1.57 &          &      \\
12 & 0.484196 & 0.417574 & 4.20 & 0.411524 & 2.53 & 0.410744 & 1.98 &
 0.410555 & 2.08 & 0.410498 & 1.85 \\
14 & 0.458965 & 0.413941 & 4.50 & 0.410954 & 2.64 & 0.410590 & 2.32 &
 0.410501 & 2.38 &          &      \\
16 & 0.442796 & 0.412240 & 4.71 & 0.410714 & 2.77 & 0.410523 & 2.69 &
          &      &          &      \\
18 & 0.432221 & 0.411414 & 4.87 & 0.410600 & 2.94 &          &      &
          &      &          &      \\
20 & 0.425210 & 0.410996 & 4.99 &          &      &          &      &
          &      &          &      \\
22 & 0.420515 &          &      &          &      &          &      &
          &      &          &      \\
\end{tabular}
\end{table}

\begin{table}
\caption{The Shanks extrapolations $A_N^{(k)}$ for the ground state
energy per site $e(N)=E_0/N$
with $k=1$ to 5 (with the VBS parameter $\alpha=1$).
The estimated decay lengths $\xi$ are obtained by applying the formula
(\protect\ref{xidef}) on the $A^{(k-1)}$.}
\label{energy}
\begin{tabular}{r|c|cc|cc|cc|cc|cc}
$N$ & $-e(N)$ & $A^{(1)}$ & $\xi$ &$A^{(2)}$ & $\xi$ &$A^{(3)}$ & $\xi$
&$A^{(4)}$ & $\xi$ &$A^{(5)}$ & $\xi$ \\
\hline
 2 & 2.000000 &          &      &          &      &          &      &
          &      &          &      \\
 4 & 1.500000 & 1.426917 & 0.97 &          &      &          &      &
          &      &          &      \\
 6 & 1.436237 & 1.408933 & 1.66 & 1.403743 & 1.50 &          &      &
          &      &          &      \\
 8 & 1.417120 & 1.404208 & 2.20 & 1.402154 & 1.86 & 1.401683 & 1.56 &
          &      &          &      \\
10 & 1.409413 & 1.402598 & 2.64 & 1.401712 & 2.11 & 1.401544 & 1.75 &
 1.401503 & 1.59 &          &      \\
12 & 1.405796 & 1.401974 & 3.00 & 1.401571 & 2.30 & 1.401505 & 1.95 &
 1.401490 & 1.83 & 1.401486 & 1.68 \\
14 & 1.403938 & 1.401713 & 3.29 & 1.401521 & 2.47 & 1.401492 & 2.17 &
 1.401486 & 2.01 &          &      \\
16 & 1.402925 & 1.401596 & 3.53 & 1.401500 & 2.64 & 1.401487 & 2.38 &
          &      &          &      \\
18 & 1.402351 & 1.401541 & 3.73 & 1.401492 & 2.82 &          &      &
          &      &          &      \\
20 & 1.402015 & 1.401514 & 3.89 &          &      &          &      &
          &      &          &      \\
22 & 1.401814 &          &      &          &      &          &      &
          &      &          &      \\
\end{tabular}
\end{table}

\begin{table} \squeezetable
\caption{The correlation functions $C_N(r)$, calculated by exact
diagonalization for $N$ up to 22.
For $N\leq 18$, these results have been
published by other authors \protect\cite{parkinson,moreo,lin}}
\label{corr}
\begin{tabular}{rccccccccc}
r & $N=6$ & 8 & 10 & 12 & 14 & 16 & 18 & 20 & 22 \\
\hline
1 & 0.47874573  & 0.47237317  & 0.46980432  & 0.46859878  &
0.46797936  & 0.46764181  & 0.46745022  & 0.46733819  & 0.46727118  \\
2 & 0.28844542  & 0.27210249  & 0.26392567  & 0.25918542  &
0.25622175  & 0.25429251  & 0.25300867  & 0.25214355  & 0.25155626  \\
3 & 0.28606604  & 0.24086913  & 0.22135314  & 0.21075706  &
0.20436261  & 0.20028789  & 0.19761348  & 0.19582835  & 0.19462472  \\
4 &  & 0.21561295  & 0.18479849  & 0.16814782  & 0.15810415  &
0.15169940  & 0.14749166  & 0.14468055  & 0.14278366  \\
5 &  & & 0.18180007  & 0.15402811  & 0.13824755  & 0.12849332  &
0.12220081  & 0.11804519  & 0.11526281  \\
6 &  & & & 0.14543474  & 0.12353893  & 0.11017252  & 0.10161828  &
0.09599964  & 0.09225227  \\
7 &  & & & & 0.12121726  & 0.10228686  & 0.09052395  & 0.08293057  &
0.07792178  \\
8 &  & & & & & 0.09842421  & 0.08305179  & 0.07322701  & 0.06679153  \\
9 &  & & & & & & 0.08143053  & 0.06847270  & 0.06012605  \\
10 &  & & & & & & & 0.06646187  & 0.05588756  \\
11 &  & & & & & & & & 0.05479614  \\
\end{tabular}
\end{table}
\end{document}